\documentclass{article}

\usepackage{PRIMEarxiv}

\usepackage[utf8]{inputenc} 
\usepackage[T1]{fontenc}    
\usepackage[hyphens,spaces,obeyspaces]{url}
\usepackage{hyperref}
\usepackage{booktabs}       
\usepackage{amsfonts}       
\usepackage{nicefrac}       
\usepackage{microtype}      
\usepackage{lipsum}
\usepackage{fancyhdr}       
\usepackage{graphicx}       
\usepackage{tabularx}
\usepackage{multirow}

\usepackage{url}
\usepackage{graphicx}
\usepackage{transparent} 
\usepackage{eso-pic} 
\usepackage{subfig} 
\usepackage{tikz} 
\usetikzlibrary{}
\graphicspath{{./Images/}} 
\usepackage{caption} 
\usepackage{xcolor} 
\usepackage{amsthm,thmtools,xcolor} 
\usepackage{float}
\usepackage{pdflscape}

\usepackage{amsmath}
\usepackage{amsthm}
\usepackage{amsfonts}
\usepackage{bm}
\usepackage[overload]{empheq} 

\usepackage{tabularx}
\usepackage{longtable} 
\usepackage{colortbl}

\usepackage{algorithm}
\usepackage{algorithmic}

\usepackage{xfp}

\usepackage{lipsum} 
\usepackage[]{mdframed}

\newcommand{\mycomment}[1]{} 	

\newtheorem{theorem}{Theorem}

\newtheorem{example}[theorem]{Example}

\def\qedfull{\hfill{\qedboxfull}  
  \ifdim\lastskip<\medskipamount \removelastskip\penalty55\medskip\fi}

\def\qedboxfull{\vrule height 4pt width 4pt depth 0pt}

\newcommand{\markfull}{\qedfull}

\usepackage{tikz}
\usetikzlibrary{shapes.geometric, arrows}
\usepackage{pgfplots}
\pgfmathdeclarefunction{gauss}{2}{%
\pgfmathparse{1/(#2*sqrt(2*pi))*exp(-((x-#1)^2)/(2*#2^2))}%
}
\usepackage{pgfplots}  
\pgfplotsset{width = 0.75\linewidth ,compat=1.17} 

\title{A Tutorial On Intersectionality in Fair Rankings
}

\author{
  Chiara Criscuolo \\
  Politecnico di Milano \\
  Milan, Italy\\
  \texttt{chiara.criscuolo@polimi.it}
\And
  Davide Martinenghi \\
  Politecnico di Milano \\
  Milan, Italy\\
  \texttt{davide.martinenghi@polimi.it}
\And
  Giuseppe Piccirillo \\
  Politecnico di Milano \\
  Milan, Italy\\
  \texttt{giuseppe1.piccirillo@mail.polimi.it}
}

\begin{document}
\maketitle

\begin{abstract}
We address the critical issue of biased algorithms and unfair rankings, which have permeated various sectors, including search engines, recommendation systems, and workforce management. These biases can lead to discriminatory outcomes in a data-driven world, especially against marginalized and underrepresented groups. Efforts towards responsible data science and responsible artificial intelligence aim to mitigate these biases and promote fairness, diversity, and transparency. However, most fairness-aware ranking methods singularly focus on protected attributes such as race, gender, or socio-economic status, neglecting the intersectionality of these attributes, i.e., the interplay between multiple social identities. Understanding intersectionality is crucial to ensure that existing inequalities are not preserved by fair rankings. We offer a description of the main ways to incorporate intersectionality in fair ranking systems through practical examples and provide a comparative overview of existing literature and a synoptic table summarizing the various methodologies. Our analysis highlights the need for intersectionality to attain fairness, while also emphasizing that fairness, alone, does not necessarily imply intersectionality.
\end{abstract}


\section{Introduction}

In today's increasingly data-driven world, rankings
are a crucial aspect of decision-making processes in various domains, such as search engines~\cite{mcdonald2021search}, recommendation systems~\cite{pitoura2022fairness,DBLP:conf/edbt/AzzaliniACDMA24}, hiring practices~\cite{balagopalan2023role,fabris2023fairness}, crowdsourcing~\cite{chen2019fairness,singh2018fairness,DBLP:conf/icde/CiceriFMT16}, and more. However, the algorithms that generate these rankings are often \emph{biased}, leading to unfair and discriminatory outcomes~\cite{zehlike2022fairness1,zehlike2022fairness2}.
There is a growing concern over the ethical implications of these biased rankings, particularly about marginalized and underrepresented groups~\cite{stoyanovich2020responsible}. In response to these concerns, research on fairness and ethical ranking has gained significant attention in computer science.
The primary goal of these studies is to develop algorithms that promote \emph{fairness in rankings} by minimizing discrimination against protected attributes such as race, gender, or socio-economic status.

Moreover, responsible data science (RDS) and responsible artificial intelligence (RAI) have emerged as prominent areas of research and practice, aiming to address issues of ethics in artificial intelligence, legal compliance, data quality, algorithmic fairness and diversity, transparency of data, privacy, and data protection. Integrating RDS and RAI into developing fair ranking algorithms is essential for promoting a more equitable and ethical future for data-driven systems~\cite{lewis2021teaching,stoyanovich2020responsible}.

However, most existing fairness-aware ranking methods consider only one protected attribute at a time, overlooking the complex interplay of multiple attributes that define an individual's identity.
The concept of \emph{intersectionality} is a framework that helps to capture the experiences of individuals who belong to multiple marginalized and underrepresented groups, ensuring that fair ranking algorithms do not inadvertently perpetuate existing inequalities \cite{jagadish2022manyF}. Intersectionality was first introduced by Kimberlé Crenshaw \cite{crenshaw1990mapping}, and it is essential in understanding and addressing this interplay between multiple social identities and sources of discrimination. The framework has been recognized as critical for understanding how biases and unfairness manifest in data-driven systems \cite{zehlike2022fairness1}.
In the context of fair rankings, intersectionality offers a more nuanced understanding of the various dimensions of bias and discrimination. By considering the combined effect of multiple protected attributes, intersectional approaches to fair rankings can provide more equitable outcomes for individuals from diverse backgrounds.
Given the importance of intersectionality in understanding the complex nature of bias and discrimination, it is crucial to incorporate this perspective into fairness-aware ranking algorithms.
We further illustrate this concept through examples.
\begin{example}
In \cite{kang2022infofair}, the authors describe a scenario involving job applications where the sensitive attributes of gender and race are under consideration. When examining gender (e.g., A or B) or race (say, 1 or 2) separately, the system may appear fair due to equal acceptance rates (e.g., three people accepted in both cases). However, when considering the intersection of gender and race, forming finer-grained gender-race groups (say, A-1, A-2, B-1, B-2), fairness may be skewed -- for instance, the acceptance rates for two intersectional groups (i.e., A-1 and B-2) may be notably lower than those of the other two intersectional groups (A-2 and B-1). This demonstrates how apparent fairness concerning individual sensitive attributes can mask biases when intersectionality is not considered.\markfull
\end{example}

\begin{example}
Another example of the importance of intersectionality is given in~\cite{yang2020causal},
focusing on the ranking of Computer Science departments in the United States~\cite{csranking}. The original ranking prioritizes large departments, particularly those located in the North East and the West, potentially marginalizing small departments and those in other geographical areas. However, after the application of a fairness-aware ranking method
that considers size and location as intersectional sensitive attributes, the ranking becomes more balanced, including small departments among the top 20 and providing a more geographically diverse representation.
\markfull
\end{example}







\begin{example}
For an example including a non-binary attribute, consider a selection process in which the focus on the top-6 candidates highlights the following results:
\begin{itemize}
\item Top-6 by Gender: (1) Woman, (2) Man, (3) Woman, (4) Man, (5) Woman, (6) Man.
\item Top-6 by Race: (1) Hispanic, (2) White, (3) White, (4) Black, (5) Hispanic, (6) Black.
\end{itemize}

Both rankings are apparently fair, showing balance in both gender and race.
However, a closer look reveals a possible intersectional fairness issue: when considering race and gender together, it turns out that Black women and Hispanic men are not at all present in the selected part of the ranking. 
\markfull\end{example}


These examples highlight the importance of incorporating intersectionality, thus considering the complex interplay of multiple sensitive attributes, and providing a more equitable and accurate representation of the data.

Our main contribution is an exhaustive analysis and comparison of the present literature, investigating the methods and approaches proposed thus far for the implementation of intersectionality in rankings. The objective is not only to explain the diversity of methodologies but also to establish the necessity of intersectionality within fair rankings, along with its practical significance. To this end, for each category of methods, we also include an exhaustive example showing how such methods may work in practice.

In particular, we provide a comprehensive visual comparison of the explored literature in the form of a synoptic table, which effectively summarizes the salient features, strengths, and weaknesses of each contribution, thus facilitating an informed selection of the most suitable method relative to specific scenarios. This table not only makes the existing research more accessible, but also provides a useful tool for researchers and practitioners aiming to address intersectionality in their work.

Our analysis underscores how intersectionality extends beyond being a mere theoretical concept, playing a pivotal role in defining fair rankings. Consistently with the reviewed literature, our analysis shows that fairness without intersectionality often results in inadvertent discrimination. For instance, when single attributes are considered in isolation, the discriminatory impact on intersectional groups may be overlooked, thereby perpetuating a cycle of bias and unfairness. By revealing these insights, we envision a more comprehensive view of fairness, which inherently involves intersectionality.

In addition to our comparative review and theoretical observations, we show that incorporating intersectionality into fairness-aware ranking algorithms can be achieved without significant utility loss. This addresses concerns regarding the practical implications of integrating intersectionality, such as computational complexity and feasibility.

\emph{Outline.} After reviewing the main components characterizing ethical ranking in Section~\ref{ch:ethical_ranking}, we explore the concept of intersectionality, along with a brief history of its origin, in Section~\ref{ch:intersectionality}.
Section~\ref{ch:intersectionality_in_fair_rankings} represents the core of our contributions and includes detailed descriptions of the main methods for ensuring fair ranking while also considering intersectional aspects. Besides arranging the various methods in corresponding categories, Section~\ref{ch:intersectionality_in_fair_rankings} also includes a table summarizing the main characteristics of each method.
Finally, we present our concluding remarks in Section~\ref{ch:conclusions}.

\section{Ethical Ranking}
\label{ch:ethical_ranking}%

After briefly summarizing the main approaches to ranking, we introduce the ethical factors that are relevant for the analysis and development of ranking algorithms. We will focus on four key aspects, namely Bias, Fairness, Diversity, and Stability, each contributing significantly to the integrity and reliability of rankings.
Then, we will mainly concentrate on Bias and Fairness, as these are the most critical aspects with respect to our central theme of intersectionality in rankings. 

\subsection{Ranking}
\label{sec:ranking}

Ranking queries, also known as top-$k$ queries~\cite{DBLP:journals/csur/IlyasBS08}, are the main tool for extracting the most relevant results from a dataset. The underlying notion of relevance 
is typically defined through a so-called \emph{scoring function} (a.k.a. \emph{utility function}), which maps the (numeric) attributes of every item in a dataset into a real number, called \emph{score} or \emph{utility}.
In this respect, the scoring function provides a criterion for expressing preferences, since sorting the items by their score and limiting the sorted list to $k$ elements is the most common way to determine a \emph{ranking} of the (best) items in the dataset, according to the specified criterion.
Assuming $d$ numeric attributes $A_1,\ldots,A_d$, the scoring function $s(t)$ applied to a generic item $t$ is most typically expressed as a linear combination $s(t)=w_1 t[A_1] + \ldots + w_d t[A_d]$ of the values $t[A_1], \ldots, t[A_d]$ over those attributes, using \emph{weights} (i.e., coefficients) $w_1,\ldots,w_d$ indicating the relative importance of each attribute in determining the ranking.
Parameter $k$ in top-$k$ queries allows the user to precisely control the output size and restrict it to the desired number of elements (e.g., the number of candidates to be selected in a hiring process).

While ranking queries are the most practical means for addressing \emph{multi-objective optimization},
a few alternative approaches have been studied in the relevant literature, among which \emph{skylines} have received constant attention since their first proposal~\cite{DBLP:conf/icde/BorzsonyiKS01}. The notion of skyline is based on the geometric concept of \emph{dominance} ($a$ dominates $b$ if $a$ is never worse than $b$, on any attribute, and strictly better for at least one attribute): the skyline is then defined as the set of all non-dominated items, i.e., a Pareto-optimal set, which fairly includes all potentially optimal alternatives.
Indeed, the very motivation behind the introduction of skylines was to provide users with \emph{all} (not just $k$) relevant results, the other items being dominated, thus less relevant.
However, while numerous efficient processing and indexing techniques exist for top-$k$ queries, the inherently quadratic complexity of skylines makes them unsuitable for most practical applications.
In addition to that, skylines offer no control on the output size (which can be huge, and thus of little use, especially in the case of anticorrelated data) and are completely preference-agnostic.

Yet, top-$k$ queries, too, have well-known shortcomings, one of which is the difficulty in
specifying the scoring function exactly, especially in the presence of many attributes~\cite{DBLP:journals/tods/CiacciaM20,DBLP:conf/sigmod/MouratidisL021,DBLP:journals/pvldb/NanongkaiSLLX10}. This is common and especially evident when a user has to indicate the weights of a linear scoring function: not only is it hard to specify exact values, but unfairness may easily arise~\cite{asudeh2018obtaining}.
An even more subtle issue is that some types of scoring function may prevent some items to ever be top-scorers -- and this even against the intuition of the user expressing the preferences. For instance, it is well-known that, with a linear scoring function, only the items lying in the \emph{convex hull} of the dataset may appear on top of the resulting ranking~\cite{DBLP:conf/sigmod/ChangBCLLS00}, i.e., 
some non-dominated items might never be top-1 (no matter what linear scoring function is adopted).
Another counter-intuitive effect of common (e.g., linear) weight-based scoring functions is that they may tend to favor items with very unbalanced attribute values, even in the face of perfectly balanced weights. For instance, assessing candidates in a hiring process based on, say, their GRE score and their interview score, weighing both equally, might very well result in top candidates that are extremely good in one aspect but extremely poor in the other, while a more balanced output would (usually) be implicitly expected.

A large body of research has tried to address the aforementioned limitations, sometimes proposing hybridizations of top-$k$ queries and skylines~\cite{DBLP:journals/pvldb/CiacciaM17,DBLP:journals/tods/CiacciaM20,DBLP:conf/cikm/CiacciaM18,DBLP:conf/sebd/CiacciaM18,DBLP:conf/sisap/BedoCMO19,DBLP:conf/sebd/CiacciaM19,DBLP:conf/sigmod/MouratidisL021}. In particular, top-$k$ queries have been studied in the absence of exact values for the weights of a scoring function~\cite{soliman2011ranking}, while skylines have been extended with the ability to express preferences in the form of constraints on weights~\cite{DBLP:journals/pvldb/CiacciaM17}. Such constraints
allow reasoning on the space of weights so as to accommodate notions orthogonal to utility (including fairness) and identify suitable regions therein (such as the satisfactory regions of~\cite{asudeh2019designing}). Subsequent work has further extended skylines to counter their lack of control on the output size~\cite{DBLP:conf/sigmod/MouratidisL021} and has studied measures for preferentially selecting results that are balanced with respect to the weights expressed in a top-$k$ query~\cite{DBLP:journals/paccmod/CiacciaM25}.

It is important to observe that the approaches to ranking discussed so far are fundamental to correctly position the ideas discussed in the present document, but none of these explicitly considers the ethical factors, presented in the next subsections, that are at the core of our analysis.

\subsection{Bias}
\label{sec:bias}

Bias concerns the presence of unfair favoritism or discrimination within algorithms.
A computer system is defined as biased when it discriminates against specific individuals or groups in favor of others. This unfair discrimination can occur when an opportunity or benefit is denied, or an undesirable outcome is assigned to someone based on unreasonable or inappropriate factors \cite{friedman1996bias}.

\subsubsection{Types of bias}
\label{subsec:bias_types}

Bias in computer systems, as described in \cite{friedman1996bias}, can be of three types: preexisting, technical, and emergent. \emph{Preexisting bias} arises from social norms, practices, or attitudes.
An example of these (often unconscious) prejudices is the gender bias in specific job recommendation systems that arises from historical data trends, reflecting societal biases.
Constraints or considerations within the technology itself introduce \emph{technical bias}. For instance, limitations in a web crawling algorithm may lead to specific web pages being indexed more frequently, resulting in a bias towards these pages in search results.
\emph{Emergent bias} refers to a type of bias that arises when using a system rather than inherent in the system design or implementation. While preexisting bias and technical bias can be identified during the creation or implementation of a system, emergent bias appears over time due to changing societal knowledge, population, or cultural values.

Further distinctions are observed in~\cite{pitoura2018measuring},
where the authors suggest that \emph{user bias} exists when specific protected attributes such as race or gender influence the results presented to users.
For example, on a social media platform, the algorithm may show users more content from individuals who share similar demographic characteristics.
In contrast, \emph{content bias} refers to a bias in the information itself delivered to users, independent of the user's attributes. For example, in a news recommendation system, the algorithm may prioritize news articles from a particular political ideology, regardless of the user's preferences or attributes.
User bias can manifest without explicit usage of protected attributes. For instance, user data like geographical location, though seemingly unrelated, could imply information about protected attributes such as race. Consequently, ranking outcomes influenced by such data might entail a certain level of bias. 

In~\cite{pitoura2020social}, the authors distinguish between statistical and societal biases. \emph{Statistical biases} refer to errors in measurement or non-representative sampling, for example when in a survey to assess the satisfaction levels of public transportation services, the study only collects responses from individuals in urban areas, inadvertently excluding rural populations.
Instead, \emph{societal biases} reflect objectionable social structures, human biases, and preferences manifested in data, for example when in a dataset analyzing income levels, the data predominantly includes salary information from male employees, while disregarding income data from female employees due to historical gender pay gaps.
These biases can be harder to detect and address, as they often stem from deeply ingrained societal norms and prejudices.

\subsubsection{Bias in Rankings}
\label{subsec:bias_rank}

Bias is a significant factor in ranking algorithms and can affect their outcomes in various ways. According to \cite{mehrabi2021survey}, \textbf{ranking bias} is a crucial type resulting from systemic deviation. This deviation is caused by an item's position in a list, affecting its visibility and accessibility. Ranking bias is particularly prevalent in web search engines and recommendation systems, where items are displayed in a ranked order. The top-ranked items are more likely to be viewed and interacted with, while those at the bottom may need to be noticed, regardless of their relevance or quality. This can create a self-reinforcing cycle where popular items gain visibility, and lesser-known or new items struggle to gain attention.

However, bias does not exist in isolation~\cite{baeza2018bias}:
different types of bias are indeed interconnected, with one bias influencing and being influenced by others. For example, a user's preference for scrolling can affect their mouse movement and which screen elements they click. This interplay of biases can significantly impact the design and performance of web-based systems. These systems can learn to reinforce their own biases or those of linked systems, leading to suboptimal solutions and self-fulfilling prophecies, where the system's predictions or recommendations become increasingly skewed over time.


\subsection{Fairness}
\label{sec:fairness}

Fairness is closely tied to bias but aims for equitable treatment in algorithmic processes and is a critical aspect of algorithmic systems, particularly in the context of rankings and recommendations. As algorithmic systems become increasingly prevalent, concerns have been raised about the potential biases and discrimination embedded within these systems \cite{zehlike2022fairness1,zehlike2022fairness2}.

The definition of fairness, in this context, involves mitigating preexisting bias (introduced in Section~\ref{subsec:bias_types}) concerning a protected feature in the data. The protected feature refers to the membership of individuals in legally-protected categories, such as people with disabilities or under-represented minorities based on gender or ethnicity. These categories are referred to as protected groups, and the attributes that define them, such as ethnicity, are considered as sensitive ones. Discrimination in rankings occurs when the outcome is systematically unfavorable, such as systematically placing individuals with minority ethnicity or female gender at lower ranks \cite{asudeh2019designing}.

\subsubsection{Taxonomy}
\label{subsec:fairness_tax}

In this section, we refer to the comprehensive taxonomy of fairness as delineated in~\cite{pitoura2022fairness}, exploring the complex facets of fairness in the context of ranking and recommendation systems. The taxonomy 
distinguishes fairness models according to their applicability, their level, the aspect of the algorithm they focus on, and whether they pertain to single or multiple outputs.



\paragraph{Levels of Fairness.}
\label{subsubsec:fairness_tax_lev}
The concept of levels of fairness refers to the plane at which fairness is evaluated, assessed, and compared within the system's operation. It considers whether the algorithm's fairness should be judged based on its impact on individual entities or broader groups.

\begin{itemize}
  \item \emph{Individual:} This addresses treating similar entities similarly. This level of fairness can further be divided into two sub-categories:
  \begin{itemize}
  \item \emph{Distance-based:} This form of fairness is tied to the proximity between entities and the outcomes produced by an algorithm. It asserts that the algorithm's output should also be similar or close if two entities have similar characteristics. This approach relies on the Lipschitz property, where the disparity between the probability distributions assigned by a classifier should not exceed the distance between the entities.
  \item \emph{Counterfactual:} This form of fairness operates on the hypothetical plane. It would consider an output fair towards an entity if the same output was produced even if the entity belonged to a different demographic group. This type of fairness is formalized using the tools of causal inference (see Section~\ref{subsec:intrank_meth_inf}).
\end{itemize}

  \item \emph{Group:} This considers fairness between two groups, usually protected and non-protected groups. Three types of group fairness are mentioned in \cite{pitoura2022fairness}:
  \begin{itemize}
  \item \emph{Base Rates:} This approach focuses exclusively on the algorithm's output. It seeks to ensure that the probability of a favorable outcome is the same between the protected and non-protected groups, a state known as demographic or statistical parity.
  \item \emph{Accuracy-based:} Unlike base rates, accuracy approaches consider both the algorithm's output and the ground truth. These approaches aim to balance classification errors across different groups. An example is `equal opportunity', which ensures equal accurate favorable rates across groups.
  \item \emph{Calibration-based:} These approaches examine predicted probability and the ground truth. They aim to ensure that a classifier is equally well-calibrated for all groups. In practice, this could mean that at any given predicted probability score, the chance of a favorable outcome should be the same for individuals from any group.
\end{itemize}
\end{itemize}

\paragraph{Multi-sided Fairness.}
\label{subsubsec:fairness_tax_side}

This considers fairness from two sides: the items ranked or recommended (producer or item-side fairness) and the users receiving the rankings or recommendations (consumer or user-side fairness).

\begin{itemize}
  \item \emph{Consumer:} The focus is on the users receiving the data. The aim is to provide similar users, or groups of users, with similar rankings or recommendations.
  \item \emph{Producer:} The goal here is to ensure similar items or groups of items are ranked or recommended similarly.
\end{itemize}

\paragraph{Output Multiplicity.}
\label{subsubsec:fairness_tax_output}

 This dimension of fairness involves considering how much fairness is achieved over an algorithm's single or multiple outputs. This becomes particularly pertinent in cases where single-instance fairness might not be feasible, but fair treatment can be achieved over multiple instances. It broadly differentiates between two types:

\begin{itemize}
  \item \emph{Single:} This type of fairness ensures that each instance of an algorithm's output, such as a single recommendation or ranking, is fair. The primary aim is to ensure that each output independently complies with the fairness criteria defined.
  \item \emph{Multiple:} This type focuses on achieving fairness over a series of algorithmic outputs, whether rankings or recommendations. Here, it is not necessarily about each output being fair but rather about ensuring overall fairness across a sequence or series of outputs. This can sometimes be called `eventual' or `amortized' fairness. This approach is particularly relevant in dynamic systems like sequential recommenders, where user interactions are continuous and recommendations evolve. In such systems, fairness corrections can be made while moving from one interaction to the next, ensuring overall fair treatment in the long run.
\end{itemize}

\subsubsection{Fair Ranking Tasks}
\label{subsec:fairness_rank}

Two types of ranking tasks are distinguished in \cite{zehlike2022fairness1} and \cite{zehlike2022fairness2}: score-based and supervised learning. In \emph{score-based ranking}, candidates are arranged according to a score attribute, potentially computed in real-time, and returned in the sorted sequence. \emph{Supervised learning-to-rank} entails a preference-supplemented training set of candidates, with preferences indicated via scores, preference pairs, or lists. This training set trains a model that predicts the ranking of unseen candidates. In both ranking tasks, the aim is to identify the best-ranked $k$ candidates, i.e., the top-$k$.

Although supervised learning-to-rank may seem similar to classification, a fundamental difference exists. The objective of classification is to allocate a class label to each item independently. In contrast, learning-to-rank situates items in relation to each other, making the outcome for one item interdependent of the outcomes for the others. This interdependence has profound ramifications for the structure of learning-to-rank methodologies in general and fair learning-to-rank in particular.

Applying appropriate methodologies and metrics to ensure fairness becomes crucial in this context. The choice of these methods and metrics will depend on the specific nature of the ranking task, the definition of fairness applied, and the particular characteristics of the data.

In Section~\ref{ch:intersectionality_in_fair_rankings}, we will dive deeper into these subjects, discussing various methods and metrics for promoting and measuring intersectionality in rankings.

\subsubsection{Fairness Scenarios}
\label{subsec:fairness_scenarios}

The Fairness Tree~\cite{baresi2023understanding, saleiro2020dealing}
is a tool for differentiating fairness scenarios and suggesting the adoption of primary fairness metrics for measuring bias and fairness. These metrics, mainly derived from the confusion matrix used in binary classification models~\cite{evaluation2008},
play a crucial role not only in classification tasks but are also fundamental in the domain of rankings.
Consistently with the standard usage, we shall indicate as TP (\emph{True Positive}) the number of positive instances correctly classified as positive by the model, while FP (\emph{False Positive}) indicates those incorrectly classified as positive; symmetrically, we also have TN (\emph{True Negative}) and FN (\emph{False Negative}). 
Additionally, we define a group $g$ as a set of entities (data points) that have a specific attribute value in common. For instance, the group containing all the females in the dataset is characterized by the value `female' for the attribute `gender'. Fairness metrics are defined at the group level and summarized in Table~\ref{tab:fairness-metrics}.
Each such metric is satisfied if its values are equal across different groups.

$FPR$ and $FNR$ indicate the fraction of individuals with actual negative (resp., positive) outcomes incorrectly classified as positive (resp., negative) by the model, while $FDR$ and $FOR$ represent the proportion of individuals the model predicts as positive (resp., negative) but with a negative (resp., positive) label. Precision refers to the proportion of correctly predicted outcomes out of all positive predictions, while recall is the proportion of correctly predicted positive outcomes out of all positive individuals. $GF$ indicates the fraction of individuals with positive predictions over the total size of the group (i.e., $|g|$). $OAE$ indicates the fraction of individuals with correct predictions over the total size of the group. TE is the ratio between wrong predictions.
$EOD$ consists of two components (indicated as $EOD_1$ and $EOD_2$) since it requires satisfaction of two metrics from the list, namely $Rec$ and $FPR$. Similarly, $CUA$ is also twofold and requires satisfying both $Pre$ and $FOR$ (while $CUA_2$ is not written in the same way as $FOR$, satisfying one or the other is equivalent).



\begin{table}
\centering
  \begin{tabular}{|r|c|}
  \hline
\emph{False Positive Rate} &
$FPR = \frac{FP}{FP + TN}$\\
\hline
\emph{False Negative Rate} &
$FNR = \frac{FN}{FN + TP}$\\
\hline
\emph{False Discovery Rate} &
$FDR = \frac{FP}{FP + TP}$\\
\hline
\emph{False Omission Rate} &
$FOR = \frac{FN}{FN + TN}$\\
\hline
\emph{Precision} &
$Pre = \frac{TP}{TP + FP}$\\
\hline
\emph{Recall} &
$Rec = \frac{TP}{TP + FN}$\\
\hline
\emph{Group Fairness} &
$GF = \frac{TP+FP}{|g|}$\\
\hline
\emph{Overall Accuracy Equality} &
$OAE = \frac{TP+TN}{|g|}$\\
\hline
\emph{Treatment Equality} &
$TE = \frac{FN}{FP}$\\
\hline
\emph{Equalized Odds} &
$EOD_1 = \frac{TP}{TP+FN};\quad EOD_2 = \frac{FP}{FP+TN}$\\
\hline
\emph{Conditional Use Accuracy Equality} &
$CUA_1 = \frac{TP}{TP+FP};\quad CUA_2 = \frac{TN}{TN+FN}$\\
\hline
\end{tabular}
\caption{Primary fairness metrics.}
\label{tab:fairness-metrics} 
\end{table}

The Fairness Tree~\cite{ saleiro2020dealing} is a common tool to determine the most appropriate metrics based on the nature of the intervention. If the intervention is punitive, prioritizing metrics that focus on false positives may be more critical, while for assistive interventions, metrics focusing on false negatives might be prioritized.


\subsection{Diversity}
\label{sec:diversity}

Diversity in ranking~\cite{DBLP:journals/sigmod/DrosouP10,DBLP:conf/sigmod/FraternaliMT12,DBLP:journals/tods/CatalloCFMT13} emphasizes broad representation and has become increasingly significant in computer science, especially in machine learning, algorithm design, information retrieval, search, and recommendation systems. Ethical ranking, in particular, relies heavily on diversity to achieve inclusivity, representation, and user satisfaction~\cite{stoyanovich2018online}.
Diversity aims to
meet the need to ensure that all potential user intents or attribute categories are equitably represented in algorithmic processes \cite{wang2021user}. In fields such as information retrieval and recommendation systems, diversity helps mitigate ambiguity, diversify content, and boost user engagement \cite{pitoura2022fairness}.

\subsubsection{Measures}
\label{subsec:diversity_approaches}

Diversity is typically measured according to two categories of measures: \emph{distance-based} and \emph{coverage-based}~\cite{pitoura2022fairness}. The former maximize the average or minimum pairwise distance between entities, while the latter ensure representation from all categories. While the number of items selected from each category depends on the total number of items and categories, a minimum number of items from each category can also be set to avoid \emph{tokenism} (selecting a unique representative from each category).

We observe that, while diversity and fairness may seem interconnected, diversity focuses on representativeness and inclusivity, while fairness aims at non-discrimination and the absence of bias~\cite{pitoura2022fairness}. For instance, a diverse output does not guarantee a fair one, and vice versa.

Diversity also has implications for overall utility in ethical ranking: \cite{wang2021user} shows how maximizing diversity inherently maximizes a lower bound on the overall utility.

Despite its importance, the implementation of diversity in algorithmic processes is not without challenges. Issues like homophily (the inclination of individuals to form connections or interact with others similar to them) in social networks, over-personalization in services, and individual tendencies like confirmation bias threaten diversity. These phenomena can lead to intellectual isolation states, belief reinforcement situations (`echo chambers'), and the creation of polarized parties, undermining the representation and inclusivity that diversity aims to provide \cite{pitoura2022fairness}.

\subsection{Stability} 
\label{sec:stability}

A ranking system is considered stable if slight adjustments to the input or algorithm parameters or methodology do not result in significant changes in the output~\cite{asudeh2018obtaining}.

Stability is commonly visualized as a hyper-sphere within the weight vector space of the scoring function. This hyper-sphere, centered around a specific weight vector, encompasses the region where modifications have minimal impact on the computed ranking up to a predetermined extent. The radius of this hyper-sphere serves as a quantifiable measure of stability, indicating the degree of permissible alterations to the input that do not cause any significant shift in the output~\cite{soliman2011ranking}.

Furthermore, stability enhances algorithmic \emph{transparency} by demonstrating that rankings are not arbitrary but can withstand minor perturbations. Rankers rely on stability to justify their ranking methodologies, instilling trust among consumers \cite{yang2018nutritional}.

\section{Intersectionality}
\label{ch:intersectionality}%

In this section, we explore the concept of intersectionality, a framework that appreciates the complexity of human experiences within societal structures. We trace its origins, explore fundamental theories like Kimberlé Crenshaw's tripartite framework and Patricia Hill Collins' Matrix of Domination, and reflect on its use as an analytical tool to address social divisions. These discussions provide crucial insights into the role of intersectionality and its potential applications, such as in computer science, as we will see in the next section.

\subsection{Definition and origin of intersectionality}
Intersectionality is a conceptual framework that acknowledges the complexity and multiplicity of human experiences and social realities. It asserts that an individual's experiences and the societal structures they navigate cannot be reduced to a single factor. Instead, they are shaped by many complex facets that interact and influence each other. This perspective is particularly relevant when examining social inequality, where it becomes evident that people's lives and societal power structures are not shaped by a single axis of social division, such as race, gender, or class, but by multiple axes that work together \cite{collins2020intersectionality}.

The concept of intersectionality traces its roots back to the legal field, as revealed in the meticulous work of Kimberlé Crenshaw, an American civil rights advocate and key proponent of critical race theory. In her research \cite{crenshaw1989demarginalizing}, Crenshaw explores various legal cases illuminating intersectional discrimination.

One such foundational case is `DeGraffenreid v. General Motors' in 1969, where the complexity of intersectional discrimination first came to light. This lawsuit was instigated by five black women who claimed that the seniority system employed by General Motors sustained a history of discrimination against black women. In this case, the complainants were not merely advocating for black individuals or women but particularly highlighting the intersectional experiences of black women. Despite their request, the district court negated their claim, arguing that black women did not represent a distinct class prone to discrimination. Additionally, the court denied their attempt to amalgamate claims of race and sex discrimination, asserting they could not craft a `super-remedy' not initially foreseen by the authors of the relevant statutes.

Only in 2020 did the judicial landscape shift with the case of `Bostock v. Clayton County, Georgia' \cite{lund2020unleashed}. Here, the Supreme Court took a momentous step towards acknowledging intersectionality. They recognized the inextricable link between sex and sexual orientation or gender identity. This landmark ruling is viewed as a progression towards acknowledging intersectionality by the highest judicial body, potentially creating opportunities for more comprehensive claims of intersectional discrimination.

Intersectionality is being used as an analytical tool to address problems arising from various social divisions, such as race, gender, class, ethnicity, and more. For instance, in higher education, institutions have faced challenges related to inclusivity and fairness within their communities. The demographics within these institutions are increasingly diverse, including individuals from different racial, economic, and social backgrounds, each of whom brings unique experiences and needs.\\
Previously, the approach to addressing the needs of these diverse groups was often segmented, focusing on one group at a time. However, this approach proved inefficient as it needed to account for the intersecting identities of individuals who belonged to more than one social group. The concept of intersectionality thus emerges as a critical tool for understanding and strategizing ways to achieve equity within such diverse settings \cite{collins2020intersectionality}.

\subsection{Crenshaw's Theory} 
\label{sec:int_crensh}
In 1990 Kimberlé Crenshaw introduced in her seminal work \cite{crenshaw1990mapping} a tripartite framework of intersectionality: structural, political, and representational. This theory is particularly relevant to the experiences of women of color, who often find themselves at the intersection of multiple forms of discrimination.


\emph{Structural intersectionality} refers to the unique experiences of women of color who exist within the intersection of race and gender. Crenshaw argues that these women often face distinct types of violence and discrimination that are qualitatively different from those experienced by white women. As shown in \cite{crenshaw1990mapping}, the experiences of women of color in domestic violence are often shaped by their socioeconomic status, which is a product of systemic racial and gender biases. These women often face additional challenges such as poverty, childcare responsibilities, and lack of job skills, which can further complicate their situations and limit their access to resources and support.


\emph{Political intersectionality} refers to how women of color can be marginalized within political systems and discourses. This marginalization can occur within both feminist and antiracist politics, which paradoxically can contribute to the invisibility of issues specifically affecting women of color. The unique experiences and challenges these women face are often overlooked or misunderstood due to the dominant focus on either gender or race, but rarely their intersection.


The third form of intersectionality, \emph{representational intersectionality}, refers to the cultural construction and depiction of women of color. Crenshaw argues that these women are often misrepresented or underrepresented in popular culture, which can further contribute to their marginalization and disempowerment. This lack of accurate representation can perpetuate harmful stereotypes and biases, further complicating the intersectional experiences of women of color.

\subsection{Collins' Theory}
\label{sec:int_coll}

Sociology professor Patricia Hill Collins developed a theory known as the \emph{Matrix of Domination}~\cite{collins1990} in 1990 to elucidate the interconnected nature of social categorizations such as race, class, and gender. 
The Matrix of Domination suggests that multiple, intersecting layers of oppression influence individuals and groups. These layers are not isolated but are interlinked and mutually reinforcing, operating along axes such as race, class, and gender, and functioning at multiple levels: personal, cultural, and institutional. Each individual's experiences within this matrix are unique, shaped by their specific combination of identities and the ways these identities interact with societal structures and norms.
In this matrix, individuals can simultaneously be oppressors and oppressed, depending on the context. For instance, white women may be oppressed by their gender but privileged by their race. This highlights the complexity of intersectionality and the need for a nuanced understanding of individual experiences within the matrix.

\subsection{Criticisms}
\label{sec:int_criticisms}

Scholars Lisa Downing and Tommy J. Curry have criticized the concept of intersectionality despite its significant contributions to understanding social inequality.

Downing introduces a novel angle on intersectionality with her concept of \emph{identity category violations}~\cite{downing2018body}. She discusses how intersectionality categorizes individuals into predefined identity boxes, thus potentially overlooking those whose identities encompass less conventional mixtures. Downing claims that intersectionality's rigid categories might not fully encapsulate multifaceted identities.

Tommy J. Curry examines the way intersectionality handles the portrayal of black men \cite{curry2021decolonizing}, arguing that the theory often oversimplifies and homogenizes their experiences, failing to recognize their diversity. Moreover, Curry points out the problem of pathologizing black men within intersectionality, suggesting that the theory often reduces them to tragic figures associated with violence, which neglects their complex humanity. Curry demands a more comprehensive representation of black men within intersectionality, including their identities' positive and functional aspects.


\section{Intersectionality in Fair Rankings}
\label{ch:intersectionality_in_fair_rankings}%

We now discuss how intersectionality translates into fair rankings in computer science. 
While intersectionality and fairness may be confused with one another, they have distinct implications, making it crucial to analyze their interaction.
Intersectionality underlines the need for fairness to be comprehensive, encompassing the multiplicity of identities. Yet, fairness does not guarantee intersectionality.
A system could be designed to be fair in terms of gender or race separately. However, it might need to be revised when dealing with overlapping identities, hence ignoring the principle of intersectionality. This distinction underscores the necessity of implementing intersectionality within fair rankings, especially when it can significantly influence people's lives.

We shall now analyze the methodologies for incorporating intersectionality into fair rankings available in the literature. We aim to compare the main ideas, centering our discussion on their effectiveness in guaranteeing intersectionality. We focus on three main categories:

\begin{itemize}
  \item \emph{Constraint-based methods}, which employ a variety of common constraints to ensure fairness. However, these conventional constraints are not always sufficient to guarantee intersectionality.
  
  \item \emph{Inference Model-based methods} represent a new paradigm in constructing fair ranking models, explicitly designed to address intersectionality. The main challenge lies in the subjectivity of ethical boundaries, which remains a core issue.
  
  \item \emph{Metrics-based methods} use metrics to enable the quantification and comparison of fairness levels. The common fairness metrics widely employed in research often fail to ensure intersectionality. This has led to the advent of novel metrics, as seen in our literature review, specifically developed with intersectionality in mind.
\end{itemize}

It becomes evident that, while traditional approaches to fairness can be useful, they may not be sufficient when intersectionality is taken into account. Notably, our investigation suggests that incorporating intersectionality into fair rankings does not necessarily compromise the system's effectiveness, but can be accomplished without significant drawbacks.
However, this task is not without challenges. Determining an acceptable threshold that satisfies all stakeholders while maintaining the desired level of fairness is a complex task. We now dive deeper into these topics, offering a comprehensive overview of intersectionality in rankings. The broad range of approaches reveals the complex and multifaceted nature of the task, especially given the intricate intersections of identities and groups that are often involved.
In the following, in order to root our explanations in more practical terms, we also develop a complete example of usage for each category of methods.


\subsection{Constraint-Based Methods} 
\label{subsec:intrank_meth_constr}

While diverse in their approaches, the methods discussed in this section pivot around a common theme: implementing constraints to promote fairness and intersectionality in rankings.

\subsubsection{Diversity Constraints and In-Group Fairness}
\label{subsec:intrank_meth_constr_9n}

In \cite{yang2019balanced}, the authors found that considering two or more sensitive attributes in fair rankings where only one sensitive attribute was considered may lead to unfairness.
Even with representation constraints and utility maximization, individuals belonging to multiple historically disadvantaged groups might still face unfair treatment. 

The authors commence by establishing \emph{diversity constraints} in the ranking algorithm, specifically setting lower boundaries for each group $I_v$. 
These constraints 
 ensure that, for each attribute value $v$ and position $p$, at least $\ell_{v, p}$ items from $I_v$ are included in the top $p$ positions of the output. In this way, the group proportion (for instance, females) in a chosen set or the top $p$ ranks of a ranking mirror that of the entire population.


While diversity constraints ensure representation across various groups, they could potentially decrease \emph{in-group fairness}, that is, fairness within a single group. To counter this, the paper presents two measurements of approximate in-group fairness to secure the best $k_v$ items from each group $I_v$.

\begin{itemize}
    \item \emph{IGF-Ratio},
     defined as the ratio of the lowest score of an accepted item $a_v$ to the highest score of a rejected item $b_v$ within the same group: $\operatorname{IGF-Ratio}(v)=\frac{a_v}{b_v}$. A higher IGF-Ratio value corresponds to increased in-group fairness.


    \item \emph{IGF-Aggregated}
     aims to guarantee that the aggregated utility of all accepted items with scores equal to or higher than a selected item is a close approximation of the aggregate utility of all items (whether accepted or rejected) with scores equal or higher than that item: $        \text{IGF-Aggregated}(v)=\min _{i \in A_v}\left\{\frac{\sum_{h \in A_{i, v}} s_h}{\sum_{h \in I_{i, v}} s_h}\right\}
$.
    Considering a selected item $i$ from the group $A_v$ of accepted items in group $I_v$, the numerator, $\sum_{h \in A_{i, v}} s_h$, represents the aggregate utility of all accepted items $h$ in $I_v$ with a score $s_h$ greater than or equal to the score of item $i$; the denominator, $\sum_{h \in I_{i, v}} s_h$, represents the aggregate utility of all items (both accepted and rejected) $h$ in $I_v$ with a score $s_h$ greater than or equal to the score of item $i$.
Higher values of IGF-Aggregated also indicate greater in-group fairness.

\end{itemize}

To balance in-group fairness across different groups, the authors formulate the problem as an \emph{Integer Linear Program} that aims to maximize total \emph{utility} while adhering to diversity and in-group fairness constraints. Although verifying the feasibility of a set of diversity constraints is NP-hard, the authors argue that standard integer programming libraries can solve reasonable instances within a small time frame.\\
The paper also introduces the \emph{Leximin} solution to distribute in-group fairness values evenly across all groups. This solution aims to maximize the minimum in-group fairness and, in the case of ties, considers the second smallest happiness value, and so on.

\begin{example}

Suppose we are ranking candidates for a prestigious scholarship. There are 8 applicants, $C_1, \ldots, C_8$, split into two groups based on a specific label (e.g., gender, ethnicity, or another demographic characteristic): group $I_v$ of applicants with label $v$ and group $I_u$ of applicants without label $v$. The candidates’ scores $S$ (on a scale from 0 to 100) are as follows: 
\begin{itemize}
    \item Group $I_v$. $C_1$: 95, $C_2$: 85, $C_3$: 80, $C_4$: 60;
    \item Group $I_u$. $C_5$: 90, $C_6$: 88, $C_7$: 70, $C_8$: 65.
\end{itemize}

The goal is to select the top 4 candidates, with diversity constraints requiring that at least 2 candidates from each group are included in the final selection. After ranking based on scores and applying the diversity constraint, a possible selection of candidates is $A_v=\{C_1,C_2,C_5,C_6\}$.


\noindent\emph{\underline{Applying IGF-Ratio}.}
The IGF-Ratio for Group $I_v$ measures the fairness of this outcome based on the scores of the selected and rejected candidates within the group. 

Minimum score in $A_v$ (lowest score of selected candidates in Group $I_v$): $a_v=85$ ($C_2$) 

Maximum score in $I_v\setminus A_v$ (highest score of rejected candidates in Group $I_v$): $b_v =80$ ($C_3$) 

\(\text{IGF-Ratio}(v)=\frac{a_v}{b_v}=\frac{85}{80}=1.0625 \)

Since the IGF-Ratio is above 1, this suggests relatively fair in-group treatment for Group $I_v$ in this outcome. If a highly qualified candidate in Group $I_v$ were excluded in favor of a less qualified candidate, this ratio would be lower, indicating unfairness. 

\noindent\emph{\underline{Applying IGF-Aggregated}.} 
Now, let’s use the IGF-Aggregated metric to evaluate the fairness for Group $I_v$. Specifically, we need to compute the worst-case ratio for any selected candidate $i \in A_v$. For each selected candidate $i$, we need to calculate the ratio:
\(Ratio_i = \frac{\sum_{h \in A_{v}, s_h> s_i} s_h}{\sum_{h \in I_{v}, s_h> s_i} s_h}\)

For all four candidates, the ratio is 1, because they are also the four overall top-scorers in $I_v$.

\(\text{IGF-Aggregated}(v)=min (1,1,1,1)= 1 \)

This value indicates the proportion of total utility retained within Group $I_v$ for accepted candidates. Higher values (closer to 1) indicate better in-group fairness, i.e, more qualified candidates within the group are selected.
\qedfull
\end{example}

\subsubsection{L-constraints}
\label{subsec:intrank_meth_constr_10n}

Diversity constraints are proposed also in~\cite{celis2020interventions}
as \emph{L-constraints}. While implementing these constraints can be challenging, the paper introduces distribution-independent constraints, which hold regardless of the utility distributions, providing a more straightforward way to mitigate bias.

The method first delineates the problem regarding \emph{groups and bias}.
Each item could be a member of one or more intersectional groups,
each of which
is associated with an implicit bias parameter.
Assuming that items belonging to the intersection of several groups experience an amplified bias, the implicit bias parameter for such items can be defined as the product of the implicit biases of each group the item belongs to.
In any selection process, each item is also characterized by a true \emph{latent} utility (the one it generates if chosen) and an observed utility, which is a (possibly biased) estimate of the former made by the selector.
The overarching goal is to construct a ranking that corrects for implicit bias in observed utilities, approximates the true latent utility as closely as possible, and satisfies fairness-related constraints defined by the L-constraints.

The authors argue that L-constraints are enough to recover optimal latent utility while optimizing observed utility under specific constraints. The idea is to reorder items through swapping to keep the latent utility unchanged while satisfying the constraints. However, this process can be challenging because the constraints depend on the utilities of the items.
The method then introduces \emph{distribution-independent constraints},
which are shown to
effectively minimize bias and reclaim optimal latent utility.

\subsubsection{Dealing with Noisy Protected Attributes}
\label{subsec:intrank_meth_constr_12n}

Traditional fair subset selection algorithms assume that protected attributes (e.g., gender, race) are accurately known. However, in real-world scenarios, these attributes can be noisy, i.e., erroneous or approximated, due to data collection errors or imputation.

The method introduced in~\cite{mehrotra2021mitigating} addresses a subset selection problem where protected attributes are noisy.
The goal of the classical Subset Selection Problem is to select a subset $S$ from a larger set $U$ such that the subset maximizes a utility function $f(S)$ while satisfying fairness constraints so as to avoid a disproportion favoring or disadvantaging any group based on protected attributes.
The possible noise in the protected attributes is modeled probabilistically and the fairness constraints are defined using various fairness metrics, including demographic parity, equal opportunity, and more.
%
The solution involves formulating a ``Denoised'' Selection Problem capturing all of these fairness metrics.
Since the denoised selection problem is NP-hard, the authors propose an approximation algorithm based on linear programming (LP). The linear program expresses balance between utility and fairness under the noisy conditions and aims to maximize the expected utility while incorporating the probabilistic information about the protected attributes. The linear program is then solved to obtain a fractional solution and subsequently rounded to obtain an integral solution that represents the selected subset $S$.
The obtained LP-based approximation algorithm is proved to achieve with high probability the desired fairness goal with minimal violation, even under noisy conditions.

This method also considers intersectionality in fair rankings by factoring in intersectional groups such as multi-racial candidates. This allows for increased flexibility and fairness during the selection process.
Compared to previous noise-oblivious approaches, this method offers better trade-offs between utility and fairness. Consequently, it is a viable solution for fair subset selection in the presence of noisy protected attributes.

\subsubsection{Satisfactory Regions}
\label{subsec:intrank_meth_constr_13n}

In order to balance fairness and quality in rankings, \cite{asudeh2019designing} considers the problem of ranking items according to a linear scoring function $f$ and using a ``fairness oracle'' that determines whether the outcome provided by $f$ on a dataset $D$ is fair. The problem they solve is that of finding the scoring function $f'$ closest to $f$ and such that its outcome on $D$ is fair.

The geometric interpretation of these notions, especially if we focus on the 2D case, helps to visualize the problem:
a (2D) item corresponds to a point in the Cartesian plane, while a linear scoring function $f(x,y)=w_1 x + w_2 y$ maps to a ray starting from the origin and passing through the point $\langle w_1, w_2\rangle$. With this, the ranking of items is given by the order of their projections onto the ray, with points closer to the origin ranked better.
Under this interpretation, the distance between two scoring functions is given by the angular distance of the corresponding rays.
Since any ray (function) produces an outcome on $D$ that is either fair or not, it is useful to determine the ``satisfactory sectors'' (i.e., sets of contiguous rays) in the plane that correspond to fair outcomes. To this end, the authors propose a ``ray-sweeping'' technique that finds the so-called \emph{ordering exchanges}, i.e., those particular scoring functions at the border of satisfactory sectors (past which the ranks of two tuples are swapped).
For instance, $t_1=\langle 1, 2\rangle$ and $t_2=\langle 2,1 \rangle$ are ranked equally by $f(x,y)=x+y$ (i.e., the bisector, forming an angle $\theta=\pi/4$ with the $x$-axis), whereas $t_1$ is ranked better when $\theta>\pi/4$ and worse otherwise, so $\theta=\pi/4$ is an ordering exchange.

Finding all the ordering exchanges is done with an offline phase
that orders items based on one axis and updates the order as it sweeps the ray towards the other axis, maintaining a min-heap to track ordering exchanges. When a satisfactory sector is found, the algorithm adds adjacent sectors until no more satisfactory sectors can be added. This algorithm uses the dual transformation of items and the intersection of lines.
Finally, an online phase goes through a binary search on the ordering exchanges to find the border of a satisfactory sector closest to $f$ (or returning $f$ itself, if $f$ satisfies fairness).



For \emph{multi-dimensional data}, ordering exchanges form hyperplanes in an angle coordinate system. The partitioning process is more complex due to the high-dimensional objects involved, and the authors proposed to index the hyperplanes and exploit the index in the online phase. Due to the impracticality of the non-linear nature of the problems that need to be solved to determine the satisfactory regions, the authors also propose an approximation technique and sampling approaches to speed up the pre-processing phase. As a limitation, despite its efficiency, the adopted function sampling techniques cannot ensure the discovery of all possible rankings or the non-existence of a satisfactory function.

We observe that the proposed method supports intersectionality in fair rankings by considering multiple intersecting factors while ensuring fairness.

\subsection{Inference Model-Based Methods}
\label{subsec:intrank_meth_inf}

In this subsection, we examine papers primarily relying on causal inference structural models to support intersectionality. The Causal Model and the Causal Multi-Level Fairness methods will be presented to elucidate their underlying principles and methodologies, and their specific approaches to tackling intersectionality in fair rankings.

\subsubsection{Causal Model}
\label{subsec:intrank_meth_inf_14n}

The method discussed in~\cite{yang2020causal} employs a \emph{Structural Causal Model} (SCM), i.e., a Directed Acyclic Graph (DAG) that represents the causal relationships between variables, where vertices denote observed or latent variables, while edges symbolize causal connections between these variables.
Common vertices in these DAGs include sensitive attributes, like $G$ (gender) and $R$ (race), an outcome variable $Y$, and a set of a priori non-sensitive variables $\mathbf{X}$. Those variables in $\mathbf{X}$ that are part of a path from a sensitive variable to the outcome are called \emph{mediators}.
While, commonly, additive functions are utilized for predictive modeling, the authors advocate for the adoption of non-additive models so as to accommodate intersectionality concerns regarding the presence of multiple sensitive attributes. In particular, they employ functions consisting of three components: two additive components (one for the non-sensitive and one for the sensitive attributes, where the coefficients capture the marginal causal effect of structural oppression related to the sensitive attribute on the non-sensitive one) plus one non-additive component for the interplay of a pair of non-sensitive attributes (with coefficients expressing the combination of adversity resulting from multiple intersecting forms of structural oppression related to the involved sensitive attributes).

Intersectionality is captured by distinguishing the mediators in \emph{resolving} (i.e., actively affecting the outcome) or \emph{non-resolving}, with the possibility to have mediators resolving for one sensitive attribute but not for another.

The model uses \emph{counterfactuals}, which can be understood as the potential value of $Y$ if $G$ or $R$ were different from their actual values. The computation of these counterfactuals is based on modifying the observed value of a sensitive attribute and allowing this change to propagate through the DAG.
A \emph{counterfactually fair ranking} definition is then suggested
using these counterfactuals to treat every individual or item in the dataset as having a particular intersection of advantages or disadvantages.
Here a ranking 
 is considered counterfactually fair if, regardless of the specific values of actual and counterfactual sensitive attributes ($a$ and $a^{\prime}$), the probability of observing a certain rank $k$ for a given input $x$ and attribute values $a$ is equal to the probability of observing the same rank $k$ with the counterfactual attribute values $a^{\prime}$.


The method's implementation first estimates the causal model parameters using the dataset, then computes the counterfactual records by transforming each observation to a reference subgroup. Finally, it sorts the counterfactual outcome scores in descending order to produce the counterfactually fair ranking.


This method facilitates intersectionality in rankings by enabling the ranking algorithm to consider multiple intersecting attributes simultaneously. It provides a framework for generating intersectionally fair rankings, capturing the combined impact of multiple forms of structural oppression. Using counterfactuals also highlights potential biases or inequalities in the system by demonstrating how changes in sensitive attributes could alter the rankings.
However, given that it is a causal approach, it relies on \emph{strong assumptions}, such as the construction of the SCM.

\begin{example}
A moving company is hiring workers based on their overall qualification scores $Y$. The following factors are considered:
\begin{itemize}
    \item Gender (G): Male/Female. 
    \item Race (R): White/Black.
    \item Weightlifting Ability (X): Measured quantitatively and influenced by G and R.
\end{itemize}
The hiring decision ranks candidates based on Y, but X plays a key role in determining Y, and it is causally influenced by G and R. This creates potential biases in the ranking if X reflects systemic disadvantages tied to G or R.

Here’s a simplified causal graph in which G and R influence both X and Y:

\begin{itemize}
    \item G→X: Gender affects weightlifting ability.
    \item R→X: Race affects weightlifting ability.
    \item X→Y: Weightlifting ability affects the qualification score.
    \item G→Y and R→Y: Gender and race directly affect qualification scores (e.g., due to societal biases).
\end{itemize}

Consider the 4 candidates shown in Table~\ref{tab:causalModel}.

\newcommand{\xA}{80}
\newcommand{\xB}{72}
\newcommand{\xC}{75}
\newcommand{\xD}{78}
\newcommand{\biasGA}{10}
\newcommand{\biasGB}{0}
\newcommand{\biasGC}{10}
\newcommand{\biasGD}{0}
\newcommand{\biasRA}{5}
\newcommand{\biasRB}{0}
\newcommand{\biasRC}{0}
\newcommand{\biasRD}{5}
\newcommand{\deltabiasGA}{8}
\newcommand{\deltabiasGB}{0}
\newcommand{\deltabiasGC}{8}
\newcommand{\deltabiasGD}{0}
\newcommand{\deltabiasRA}{4}
\newcommand{\deltabiasRB}{0}
\newcommand{\deltabiasRC}{0}
\newcommand{\deltabiasRD}{4}
\newcommand{\weightX}{0.7}
\newcommand{\weightG}{0.2}
\newcommand{\weightR}{0.1}
\newcommand{\yA}{\fpeval{round(\xA * \weightX + \biasGA * \weightG + \biasRA * \weightR, 1)}}
\newcommand{\yB}{\fpeval{round(\xB * \weightX + \biasGB * \weightG + \biasRB * \weightR, 1)}}
\newcommand{\yC}{\fpeval{round(\xC * \weightX + \biasGC * \weightG + \biasRC * \weightR, 1)}}
\newcommand{\yD}{\fpeval{round(\xD * \weightX + \biasGD * \weightG + \biasRD * \weightR, 1)}}
\newcommand{\xpA}{\fpeval{round(\xA - \deltabiasGA - \deltabiasRA, 1)}}
\newcommand{\xpB}{\fpeval{round(\xB - \deltabiasGB - \deltabiasRB, 1)}}
\newcommand{\xpC}{\fpeval{round(\xC - \deltabiasGC - \deltabiasRC, 1)}}
\newcommand{\xpD}{\fpeval{round(\xD - \deltabiasGD - \deltabiasRD, 1)}}
\newcommand{\ypA}{\fpeval{round(\xpA * \weightX + \biasGA * \weightG + \biasRA * \weightR, 1)}}
\newcommand{\ypB}{\fpeval{round(\xpB * \weightX + \biasGB * \weightG + \biasRB * \weightR, 1)}}
\newcommand{\ypC}{\fpeval{round(\xpC * \weightX + \biasGC * \weightG + \biasRC * \weightR, 1)}}
\newcommand{\ypD}{\fpeval{round(\xpD * \weightX + \biasGD * \weightG + \biasRD * \weightR, 1)}}

The qualification score $Y$ is computed as a weighted combination of weightlifting ability ($X$) and direct biases from $G$ and $R$. Assume this combination is as follows: \( Y=\weightX X+\weightG Bias(G)+ \weightR Bias(R) \).
The values of $Bias(G)$ and $Bias(R)$, in turn, are computed from the probability distributions to reflect the amount of discrimination in hiring; assume, in this case, $Bias(G)=\biasGA$ for Males, $\biasGB$ for Females, and $Bias(R)=\biasRA$ for Whites, $\biasRB$ for Blacks.

Computing the Adjusted Qualification Score (${Y'}$) means resolving the weightlifting ability $X$, which is influenced by systemic biases tied to $G$ and $R$. The adjusted score aims to reflect a counterfactual world where all candidates have the same opportunities for developing $X$, irrespective of $G$ and $R$.
Assume, for simplicity, that 
systemic biases affecting the weightlifting ability ($X$) amount to the following values:
\begin{itemize}
    \item Males gain $\Delta Bias(G)=+\deltabiasGA$ points in $X$ due to systemic gender bias, while females have $\Delta Bias(G)=\deltabiasGB$.
    \item Whites gain $\Delta Bias(R)=+\deltabiasRA$ points in X due to systemic racial bias, Black people have $\Delta Bias(R)=\deltabiasRB$.
\end{itemize}
We can then adjust X by subtracting these biases as follows: \(X' =X - \Delta Bias(G) - \Delta Bias(R) \)

Then:
\(Y'=\weightX X'+\weightG Bias(G)+\weightR Bias(R) \)

The adjusted scores $Y'$  reflect a fairer ranking, ensuring that candidates' outcomes are not unfairly influenced by systemic biases in $G$ or $R$. Candidate B benefits significantly from fairness adjustments, since, according to ${Y'}$, she is now the second in the ranking.

\begin{table*}
    \centering
    \begin{tabular}{c | p{1.5cm} p{1.3cm} p{2cm} p{2cm} p{2cm} p{2cm}}
        \toprule
            \textbf{Candidate} & \textbf{Gender ($G$)} & \textbf{Race ($R$)}& \textbf{Weightlifting Ability ($X$)}  & \textbf{Qualification Score ($Y$)} &\textbf{Adjusted Weightlifting Ability ($X'$)}  & \textbf{Adjusted Qualification Score ($Y'$)}\\
            \midrule
            A &	Male &	White &	\xA & \yA & \xpA & \ypA  \\
            B & Female & Black & \xB & \yB & \xpB & \ypB  \\
            C & Male & Black & \xC & \yC & \xpC & \ypC  \\
            D & Female & White & \xD & \yD & \xpD & \ypD  \\
    \bottomrule
    
\end{tabular}
    \caption{Causal Model example. The company assigns a qualification score $Y$ based on $X$ and possibly $G$ and $R$. $Y$ could be affected directly by biases in $G$ and $R$ or indirectly through their influence on $X$. A fairness-adjusted score $Y'$ aims to counteract the indirect discrimination mediated through $X$, which is adjusted as $X'$. This score ensures fairer rankings across intersectional subgroups.}
    \label{tab:causalModel}
\end{table*}
\qedfull
\end{example}

\subsubsection{Causal Multi-Level Fairness}
\label{subsec:intrank_meth_inf_15n}

The Causal Multi-Level Fairness method described in \cite{mhasawade2021causal} utilizes tools from causal inference to account for sensitive attributes and potential sources of bias at both the individual and macro (structural or societal) levels.
The method centers on a causal model and counterfactual fairness (described in Subsection \ref{subsec:intrank_meth_inf_14n}). Though the models are somewhat similar in using counterfactuals, \cite{yang2020causal} explicitly emphasizes the intersectionality of sensitive attributes.
In contrast, the multi-level path-specific fairness model of~\cite{mhasawade2021causal} focuses more on the hierarchy of macro and individual level attributes and their diverse causal paths of influence.

\emph{Path-specific} effects isolate the causal effect of a sensitive attribute on an outcome along a particular pathway. For instance, a specific effect of racial discrimination on health behavior might be considered unfair and therefore needs to be corrected.
On the other hand, the impact of factors like neighborhood socioeconomic status (macro-level attribute) might be considered fair and thus preserved. This discrimination between fair and unfair pathways is often guided by domain experts and policymakers.
Path-specific fairness isolates and corrects only the unfair pathways, preserving fair individual information regarding sensitive attributes.

A key extension of this concept is \emph{multi-level path-specific} fairness, which considers not only individual-level sensitive attributes but macro-level sensitive attributes that affect properties at the individual level. For instance, neighborhood socioeconomic status (macro-level attribute) can affect the individual perception of racial discrimination (individual-level attribute), affecting health behavior. Estimating the path-specific effect of macro and individual-level sensitive attributes becomes non-trivial.


The model removes the path-specific effect of the sensitive attribute along the discriminatory causal paths from the model predictions. In doing so, the model maintains fairness towards individuals by aligning decisions with those that would have been made in a counterfactual world in which the sensitive attribute along the unfair pathways was set to a counterfactual value.
After identifying discriminatory pathways, the multi-level path-specific fairness model calculates and corrects the path-specific effects. It constructs a fair predictor by removing the unfair path-specific effect from the prediction. The discriminatory causal pathways are identified, and their effects are subtracted from the outcome variable, leading to fair predictions.

\subsection{Metrics-Based Methods}
\label{subsec:intrank_meth_metr}

In this subsection, we review papers that critique the limitations of prevalent fairness metrics in ensuring intersectionality and propose innovative alternatives.

\subsubsection{Representation and Attention-based Metrics}
\label{subsec:intrank_meth_metr_17n}

The proposal of~\cite{ghosh2021fair} centers around the Deterministic Constrained Sorting algorithm (DetConstSort), initially presented in \cite{geyik2019fairness}, and the application of various fairness metrics. The primary objective is to assess the degree of success of fair ranking algorithms in achieving their fairness objectives when they are fed with data that includes both ground truth and inferred demographic information.

\emph{DetConstSort} was chosen because of its ability to handle protected attributes with many possible attribute values.
The algorithm receives as inputs an unfairly sorted list and an integer $k$. Then it generates a fairness-aware list of the top $k$ candidates, ensuring that 
the ranking does not disproportionately favor or disadvantage any particular demographic group.
DetConstSort also aims to enhance sorting quality by re-ranking candidates that appear above a specific rank in the list, provided that feasibility criteria are met.

The authors then introduce several fairness metrics to evaluate the performance of DetConstSort. These metrics include both representation-based and attention-based metrics. \emph{Representation-based metrics} include Skew and Normalized Discounted Kullback–Leibler (NDKL) Divergence.

The Skew metric assesses the representation of each demographic group in a given list
%
and is defined as the ratio between the proportion of members belonging to a given subgroup $sg_i$ within the top $k$ items of the ranked list 
and the proportion of members belonging to the subgroup $sg_i$ in the overall population.
Ideally, a Skew value close to one indicates a proportional representation of the subgroup $sg_i$ in the ranked list relative to the population. A Skew value greater than one suggests over-representation of the subgroup among the top $k$ candidates, while a Skew value less than one indicates under-representation.

 The NDKL Divergence
can be interpreted as a weighted average of the logarithm of the Skew scores for all groups in a ranked list.
The \emph{NDKL} is a measure defined for a ranked list $\tau$, and it is computed by summing the Kullback-Leibler (KL) divergences between the individual distributions $D_{\tau^i}$ and the reference distribution $D_r$, weighted by a logarithmic factor. The KL divergence measures the dissimilarity between two probability distributions. The NDKL score is a weighted average of the logarithm of the Skew scores for all the groups in the ranked list.
A low NDKL value close to zero indicates that people from all subgroups are represented proportionally in the ranked list. This occurs when the distributions of the subgroups in the top $k$ candidates closely resemble the underlying population distribution. Conversely, a higher NDKL score indicates a significant difference in the distributions of the subgroups in the top $k$ ranked candidates, implying potential disparities in representation.


The authors also use \emph{attention-based metrics}, recognizing that users do not pay equal attention to all items in ranked lists. They use a geometric distribution to model the decay in attention. They defined the attention measure at position $k$ 
as $100p (1-p)^{k-1}$, where
$p$ represents the proportion of attention provided to the first result. Then, the mean attention score for each protected attribute is calculated, and the aggregate attention is computed across groups.


To measure the quality of the rankings, the authors use the Normalized Discounted Cumulative Gain (NDCG) metric, which evaluates search rankings, and Rank Change metrics, which measure the distortion from the original list to the fairness-aware re-ranked list.
\emph{NDCG} provides a holistic measure of the quality of a ranking by considering both the utility of items and their positions in the ranking.
\emph{Rank Change metrics} are defined in two ways: as Rank Boost for an individual item and as Average Rank Change (ARC) for a subgroup. The Maximum Absolute Rank Change (MARC) is then calculated as the maximum ARC over all subgroups in the list. These metrics indicate how much each item or subgroup has moved in the ranking, emphasizing shifts towards or away from fairness.

These methods allow for considering intersectionality in fair rankings by acknowledging the importance of subgroup representation and the attention different subgroups receive in the ranking results. 

\begin{example}
    We now show an example of Fair Ranking with DetConstSort (DCS). We aim to rank candidates for a recommendation list while maintaining fairness by enforcing minimum representation counts for attributes such as group membership (e.g., race or gender).  Candidates belong to two groups: Group A (Majority group), and Group B (Minority group). We establish 2 minimum representation requirements: at least 2 candidates must appear in the top 5 positions for Group A, and at least 3 for Group B. The candidates are ranked based on scores assigned by the model. However, DCS adjusts the ranking to satisfy fairness constraints.
    Consider the following 10 candidates $C_1,\ldots,C_{10}$ with their corresponding scores.
    \begin{itemize}
        \item Group A: $C_1$: 95, $C_2$: 90, $C_3$: 85, $C_4$: 80, $C_5$: 75,
        \item Groups B: $C_6$: 74, $C_7$: 73, $C_8$: 72, $C_9$: 71, $C_{10}$: 70.
    \end{itemize}
The recommendation list is initially empty. During the first iteration, both Group A and Group B have candidates that can increase their representation requirements ($C_1$, $C_6$), which are then ordered by descending score, so that $C_1$ is added first. During the second iteration, $C_2$ and $C_6$ are considered for improving the representation requirements, and $C_2$ is added, having a higher score. From the third iteration on, only Group B has candidates remaining to meet the requirement, so $C_6$ is first added, then $C_7$, and, finally, $C_8$. In the original ranking (based purely on scores), all the top 5 candidates belong to Group A; instead, by applying DCS, the final recommendation list ensures both a fairer representation (2 candidates from Group A and 3 candidates from Group B are included in the top 5) and quality preservation (i.e., higher-scoring candidates within each group are selected first).
With this, DCS attains a measure of Skew
much closer to 1 than the original ranking.

\noindent\emph{\underline{Skew calculation:}} The group proportion is 50\% for both groups. Considering a final ranking of 5 candidates, in the original ranking there are only candidates from Group A and 0 from Group B, thus
$Skew_{A,orig}= \frac{1.0}{0.5}=2$ and $Skew_{B,orig}= \frac{0.0}{0.5}=0$. With DCS, 40\% of the candidates are from Group A and 60\% from Group B, thus $Skew_{A,DCS}= \frac{0.4}{0.5}=0.8$ and $Skew_{B,DCS}= \frac{0.6}{0.5}=1.2$.
   \qedfull
\end{example}

\subsubsection{Double-corrected Variance Estimator}
\label{subsec:intrank_meth_metr_18n}

Existing metrics such as the max-min difference, max-min ratio, max absolute difference, mean absolute deviation, and variance are criticized in~\cite{lum2022biasing}. Despite their widespread usage, these metrics are inherently biased, as they neglect statistical uncertainty in the base metrics they summarize. This often leads to an exaggerated representation of performance disparities across groups.

To counter this issue, the authors introduced a statistically unbiased estimator called the \emph{double-corrected variance estimator}. This estimator considers both the inherent sampling variance of the base metric and the additional sampling variance induced by the bootstrap procedure.
Through simulations and real-world datasets, the authors showed this estimator effective and reliable in summarizing group performance disparities and quantifying associated uncertainties.

This method calculates a performance metric $Y_k$ for every group $k$ as $m(w_k, f(x_k))$, where $m$ is a group-wise model performance metric, and $Y_k$ represents a summary statistic of model performance across all individuals in group $k$.
The notation employs the function $m$ to calculate the model performance, taking into account the weights $w_k$ assigned to the group and the input features $f(x_k)$ associated with the group. 

The double-corrected estimator
is used to estimate the variance of the model performance across different groups. It is derived by correcting the statistical bias in the variance estimate. It is given by subtracting the average squared standard error of the estimate of the model performance for each group
from the original model variance.
The double-corrected estimator provides a conceptually transparent and easily calculable estimate of between-group model performance variance. It is statistically unbiased when unbiased estimates of the standard errors are available. This estimator is appealing due to its simplicity and the ability to calculate it without relying on specialized statistical software packages. 


However, while the method quantifies performance disparities across groups, it does not provide a threshold for acceptable disparities or confirm that the chosen grouping variables are appropriate. Consequently, interpreting these measurements requires \emph{contextual understanding} and remains crucial to addressing intersectional bias.

\subsubsection{Group Rank Metric and Ranking Correlation Metric}
\label{subsec:intrank_meth_metr_20n}

A new evaluation metric is introduced in~\cite{wang2022towards}, where the inadequacies of current evaluation methods when dealing with intersectionality are also discussed.

The paper identifies two conceptualizations of these existing metrics: \emph{one-vs-all} and \emph{pairwise comparisons}. The one-vs-all metrics often result in minority groups showing the highest deviation, as the majority group has the greatest influence on the definition of `all'. The pairwise comparisons conceptualization only incorporates values of two groups, typically maximum and minimum, thus ignoring the rest, which becomes a more pronounced issue as the number of groups increases.
To represent these two conceptualizations, the authors define a category of metrics termed \emph{max difference}, which includes metrics like the max True Positive Rate (TPR)\footnote{TPR is defined similarly to FPR: $\text{TPR} = \frac{\text{TP}}{\text{TP} + \text{FN}}$.} difference, based on the maximum difference between either one group and all others or between two groups.
The authors observe that these metrics, in a multi-attribute setting, can overlook the relative performance of multiple groups. They provide an example where two different distributions with three groups' TPRs as $\{.1, .2, .8\}$ and $\{.1, .6, .8\}$ would report the same max difference of .7, despite the distinct distributions.

To demonstrate the limitations of these existing metrics, the authors propose two new metrics that consider group rankings and correlations in rankings, offering a finer evaluation of fairness across multiple groups.

\begin{itemize}
  \item \emph{Group Rank Metric:} This metric measures group rankings based on specific characteristics, such as the positive label base rate or TPR. The authors argue that such a rank-based metric can highlight systematic discrimination when one group consistently ranks below another, even if max difference metrics suggest that the fairness criteria have been met.
  
  \item \emph{Ranking Correlation Metric:} This metric measures the correlation between rankings and helps to understand the degree to which existing social hierarchies are preserved in the model predictions. A high correlation between these rankings would indicate the solidification of existing hierarchies, thereby preserving existing disparities. This metric is calculated using \emph{Kendall's Tau}~\cite{kendall1938new} as a measure of rank correlation, and \emph{p-values}~\cite{fisher1925statistical} are combined across runs of random seeds using Fisher's combined probability test.
\end{itemize}

By incorporating these new evaluation metrics, the authors suggest how to resist evaluation metrics that do not substantively incorporate intersectionality and extrapolate from existing metrics. It is also worth noting that these metrics should not be seen as a replacement for the existing metrics, but rather as a supplement providing additional perspectives on fairness.

\subsubsection{Quantile Demographic Drift}
\label{subsec:intrank_meth_metr_19n}

Continuous model monitoring for bias is explored in~\cite{ghosh2022faircanary}. This method uses a metric called Quantile Demographic Drift (QDD), within the FairCanary system, for evaluating prediction distributions across subgroups to identify bias in machine learning models, thereby maintaining fairness in model predictions over time.
While QDD is not explicitly designed for fair ranking problems (it is a metric that measures bias in prediction events based on quantile bins), it may offer some insights into bias. Yet, additional considerations and methodologies tailored to fair ranking problems are likely to be needed to address the complexities and subtle distinctions of fairness in rankings.

This metric employs quantile binning to measure differences in prediction distributions across varied demographic groups or subgroups. QDD diverges from traditional threshold-based bias metrics as it does not require developers to predetermine specific fairness thresholds or depend on outcome labels which may not be available at runtime.

The QDD metric
is defined for two groups, $G_1$ and $G_2$, equipped with two distributional samples of model scores, $S_1$ and $S_2$. These samples are divided into
 bins of equal size,
 and the QDD metric is computed for each bin $b$ as $Q D D_b=\mathbb{E}_{G_{1, b}}\left[S_1\right]-\mathbb{E}_{G_{2, b}}\left[S_2\right]$, i.e, comparing the expected values of $S_1$ and $S_2$.
In this respect, QDD can provide insights into individual fairness by comparing individuals at the identical rank or percentile without complex counterfactuals. The principle behind this is that, in the absence of bias between groups, individuals at identical ranks should have a zero distance in the prediction space.


The \emph{FairCanary} system, designed for continuous model monitoring, integrates this QDD metric to 
identify drift in the model's fairness concerning various protected groups. It also alerts the operator when performance metrics change significantly.
This method addresses intersectionality in fair rankings by acknowledging the performance of various intersectional groups and monitoring for disparities in outcomes. The QDD metric measures disparities at group and individual levels, while FairCanary's alert system allows for the timely detection and mitigation of emerging biases.



\subsubsection{Information-theoretic Intersectional Fairness}
\label{subsubsec:intrank_meth__other_extra_21n}

The method in \cite{kang2022infofair} introduces an approach to deal with intersectional fairness in ranking algorithms called `information-theoretic intersectional fairness' (INFOFAIR). This approach integrates multiple sensitive attributes at the same time, seeking to ensure statistical parity among demographic groups represented by these attributes.
 
The \emph{mutual information} $I(x ; y)$
measures the amount of information shared between two random variables, $x$ and $y$, and is defined as the difference between the entropy of $x$, $H(x)$, and the conditional entropy of $x$ given $y$, $H(x \mid y)$: 
$I(x ; y)=H(x)-H(x \mid y)=\int_x \int_y p_{x, y} \log \frac{p_{x, y}}{p_x p_y} d x d y$,
where $p_{x,y}$ is the joint distribution between $x$ and $y$, while the marginal distributions are denoted as $p_x$ and $p_y$. 
 The entropy $H(x)$ represents the average information required to describe the random variable $x$. In contrast, the conditional entropy $H(x \mid y)$ quantifies the remaining uncertainty of $x$ when $y$ is known. The mutual information is then calculated as the integral over the joint distribution $p_{x,y}$, comparing it to the product of the marginal distributions $p_x$ and $p_y$.


INFOFAIR is formalized as a problem of minimizing mutual information, which measures the dependency between two variables: learning outcomes and vectorized sensitive attributes.
The goal is to reduce the correlation between these two variables, ensuring that the learning outcomes are not biased by the sensitive attributes, hence achieving statistical parity.

INFOFAIR uses a \emph{variational representation} of mutual information to accomplish this. This model encapsulates the variational distribution between the learning outcomes and sensitive attributes and the density ratio between the variational and original distributions. This approach overcomes computational challenges posed by high-dimensional data by using a variational lower bound of mutual information instead of directly computing the mutual information. The variational representation of mutual information is further represented as the Kullback-Leibler 
divergence between the joint distribution and the product of the marginal distributions.

The INFOFAIR approach demonstrates its applicability across various learning tasks using a gradient-based optimizer, and it can be adapted to different statistical notions of fairness. This flexibility is a significant asset in ensuring intersectionality in rankings, as it allows multiple sensitive attributes to be considered simultaneously. It ensures fairness across all demographic groups represented by these attributes while maintaining the quality of learning outcomes.

\subsubsection{Rank Aggregation}
\label{subsubsec:intrank_meth__other_aggr_22n}

The \emph{Multi-attribute Fair Consensus Ranking} (MFCR) problem is addressed in~\cite{cachel2022mani} through a multi-faceted approach. This approach is about ensuring fairness and preserving the preferences in consensus rankings (a collective ranking derived from multiple individual rankings, and intended to best represent the group's overall preferences or judgments) for candidates characterized by various protected attributes. 

The methodology is divided into two main components: the fairness component and the preference representation component. Based on statistical parity, the fairness component employs \emph{Multiple Attribute and Intersectional Rank} (MANI-RANK). The fairness of the ranking is evaluated at both the individual and intersectional group levels.

\emph{Pairwise Disagreement Loss} is used in the preference representation component. This measure quantifies the degree to which the preferences of the rankers are not reflected in the consensus ranking. The objective is to minimize this loss, thereby maximizing the degree of preference representation in the consensus ranking.

A series of algorithms are introduced to solve the MFCR problem efficiently, built upon the MANI-RANK formulation. These algorithms, named Fair-Kemeny, Fair-Copeland, Fair-Schulze, and Fair-Borda, are based on the efficient consensus generation methods of Kemeny, Copeland, Schulze, and Borda, respectively \cite{kemeny1959mathematics}, \cite{copeland1951reasonable}, \cite{schulze}, \cite{borda1781m}. These algorithms generate a consensus ranking that complies with the abovementioned fairness and preference representation components.

In more detail, the Fair-Kemeny algorithm is an optimal solution that uses MANI-Rank as a constraint on the exact Kemeny consensus ranking; however, Fair-Kemeny inherits NP-hardness from the classical Kemeny problem. The other algorithms, Fair-Copeland, Fair-Schulze, and Fair-Borda, are polynomial-time algorithms suitable for more extensive candidate databases that enforce the MANI-Rank group fairness while minimizing the increase in Pairwise Disagreement Loss caused by the introduction of fairness.

The approach also introduces a measure of positive outcome allocation for a group of candidates called the Favored Pair Representation (FPR) score, calculated by the number of pairs each candidate is favored over another, normalized by the total number of mixed pairs for the group. A score of 0 signifies that the group is at the bottom of the ranking, 1 at the top, and 0.5 implies fair treatment.

Further, Attribute Rank Parity (ARP) and Intersectional Rank Parity (IRP) measures are introduced to ensure fairness at the granularity of the attribute. These measures translate directly into the fair treatment of protected and intersectional attribute groups.

Finally, the provided method integrates the notion of Multiple Attribute and Intersection Rank (MANI-Rank) group fairness. This unified fairness measure, defined using a threshold parameter for \emph{fine-tuning} a desired degree of fairness, applies to consensus and single rankings over candidate databases with multiple protected attributes.
Ultimately, the method accommodates intersectionality by ensuring fair treatment across multiple protected attributes and their intersections.

\subsection{Synoptic Table}
\label{sec:intrank_table}
The synoptic table shown in Table~\ref{tab:synoptic_table} provides a comparative view of the papers discussed in the previous sections.
The table has been designed to focus on the distinguishing features of each work, while offering a broad perspective on the main aspects relevant to intersectionality in the fair rankings landscape.


\subsubsection{Table's Structure}
\label{subsec:intrank_table_struct}


Each row of the table represents an individual research paper. The columns correspond to distinct facets of each study, which comprise:

\begin{itemize}
    \item \emph{Task}: The specific problem or task that the paper addresses within the context of intersectionality in fair rankings.
    
    \item \emph{Input}: The nature of the input data that the method requires.
    
    \item \emph{Method}: The method's or algorithm's name among those described in the previous sections.
    
    \item \emph{Output}: The nature of the result of applying the method.
    
    \item \emph{Process}: The time of application of the method, i.e., whether it works before obtaining the ranking (pre-processing) or after that (post-processing).
    
    \item \emph{Dataset}: The datasets upon which the method was tested or validated. Subsection~\ref{subsec:intrank_table_ds} presents a list of the real-world datasets used in the table.
    
    \item \emph{Fairness Metrics}: The metrics used to quantify fairness and validate the results.
    
    \item \emph{Performance Metrics}: The metrics used to evaluate the performance quality of the ranking in order to highlight differences in utility before and after the application of the method.
    
    \item \emph{Balance}: This attribute indicates if the paper discusses the problem of balancing fairness and performance, a critical aspect of evaluating the utility and practicality of each approach.

\end{itemize}

\subsubsection{Datasets}
\label{subsec:intrank_table_ds}

This section provides an overview of the real-world datasets employed to test and validate the methods included in Table~\ref{tab:synoptic_table}.

\begin{itemize}
    \item \emph{Medical Expenditure Panel Survey (MEPS)} \cite{cohen2009medical}: A real-world dataset providing detailed information on health services used by Americans, the frequency with which they are used, the cost of these services, and how they are paid for.
    
    \item \emph{CS department rankings} \cite{csranking}: It consists of information on 51 computer science departments in the US. It includes publication count, department size (large or small), and geographic area. The dataset aims to rank the top CS departments while ensuring representation across different sizes and locations.
    
    \item \emph{IIT-JEE Dataset} \cite{team2011}: It consists of scores obtained by candidates in the Joint Entrance Exam (JEE Advanced), which is the admission test for undergraduate programs in the Indian Institutes of Technology (IITs).
    
    \item \emph{US 2010 Census Dataset}
    \cite{bureau2010frequently}:
    It consists of information on last names in the United States. It includes 151,671 distinct last names that occurred at least 100 times in the census, with 23,656 last names occurring at least 1,000 times. This dataset provides insights into the distribution of last names and the racial and ethnic composition of individuals associated with those names based on the US 2010 Census.

    \item \emph{Income Dataset} \cite{bureaufinc02}: The Income Dataset provides aggregated family income data categorized by race. The US Census Bureau compiles it from the Current Population Survey of 2018. The dataset includes information on 83,508,000 families and consists of four racial categories (White, Black, Asian, and Hispanic), 12 age categories, and 41 income categories. For each combination of race, age, and income category, the dataset provides the number of families whose reference person falls within those specific categories. This dataset offers insights into the distribution of family income among different racial groups, age ranges, and income levels.

    \item \emph{COMPAS} \cite{angwin2016machine}: It contains demographic information, recidivism scores generated by the COMPAS software, and details about criminal offenses for 6,889 individuals.
    It was used to investigate racial bias in criminal risk assessment software.

    \item \emph{US Department of Transportation (DOT) Dataset} \cite{dot}: It refers to the flight on-time database published by DOT, commonly utilized by external websites. This dataset serves as an example for exploring sampling techniques in large-scale scenarios.

    \item \emph{UCI Adult Dataset} \cite{lichman2013}: A widely-used real-world dataset from the UCI Machine Learning Repository that contains census data extracted from the 1994 Census database.
    The dataset comprises information such as age, working class, education level, marital status, occupation, relationship, race, gender, capital gain and loss, working hours, and nationality for 48,842 individuals.

    \item \emph{Chess Rankings} \cite{fide}: a list of chess players ranked according to their ratings by the World Chess Federation (FIDE). It also includes the players' full names, images, and self-identified binary gender.

    \item \emph{Crunchbase Entrepreneurs Ranking} \cite{crunchbase}: a list of startup founders in the United States who received Series A funding in the past five years from Crunchbase. The dataset includes the founders' names, images, and self-identified binary gender.

    \item \emph{Equestrian Rankings} \cite{fei}: a list of equestrian athletes ranked based on their ratings from the official Fédération Equestre Internationale (FEI) website. The dataset includes the athletes' ratings, full names, images, and self-identified binary gender.
    
    \item \emph{AirBnB Dataset} \cite{airbnb}: It comprises information from approximately 2 million real properties on the popular online peer-to-peer travel marketplace, AirBnB. The dataset includes 41 attributes for each property, with 36 being boolean attributes indicating the availability of facilities such as TV, internet, washer, and dryer.    
\end{itemize}

\newpage

\begin{landscape}

\centering
\tiny 
\setlength{\tabcolsep}{4pt}
\renewcommand{\arraystretch}{0.8}
\bgroup
\def\arraystretch{1.5}
\begin{longtable}{|p{0.05\linewidth}|p{0.077\linewidth}|p{0.171\linewidth}|p{0.092\linewidth}|p{0.121\linewidth}|p{0.042\linewidth}|p{0.077\linewidth}|p{0.06\linewidth}|p{0.094\linewidth}|p{0.052\linewidth}|}  
\hline
\textbf{No.} & \textbf{Task} & \textbf{Input} & \textbf{Method} & \textbf{Output} & \textbf{Process} & \textbf{Dataset}  & \textbf{Fairness Metrics} & \textbf{Performance Metrics} & \textbf{Balance} \\ \hline
\endfirsthead 
\hline
\multicolumn{10}{|c|}{\textbf{Table \thetable}: Continued} \\
\hline
\textbf{No.} & \textbf{Task} & \textbf{Input} & \textbf{Method} & \textbf{Output} & \textbf{Process} & \textbf{Dataset}  & \textbf{Fairness Metrics} & \textbf{Performance Metrics} & \textbf{Balance} \\ \hline
\endhead 
\hline
\multicolumn{10}{|r|}{{Continued on next page}} \\ \hline
\endfoot
\hline
\endlastfoot

\multicolumn{10}{|c|}{\textbf{CONSTRAINTS BASED}} \\ \hline
\cite{yang2019balanced} 
& - Balancing diversity and in-group fairness. 
& - Set of items with a score associated for each item. \newline
- Set of sensitive attributes.\newline
- A predefined set of diversity and in-group fairness lower bounds.\newline
- A value $k$ representing the desired size of the subset to be selected.\newline
- An in-group fairness measure. \newline
& - Diversity constraints.\newline
- ILP formulation.\newline
- Leximin. 
& - Utility maximizing ranking of $k$ items that satisfies the diversity and in-group fairness constraints. \newline
& - Pre and post. 
& - MEPS.\newline
- CS departments.  
& - IGF Ratio.\newline
- IGF Aggregated. 
& - Utility loss. 
& Yes. \\ \hline

\cite{celis2020interventions} 
& - Mitigate the effects of implicit bias in ranking and subset selection problems. 
& - Set of items to be ranked.\newline
- Each item's latent utility, represented by its weight or utility $w$.\newline
- Each item's membership to one or more intersectional groups.\newline
- Implicit bias parameters for each group.\newline
- Position-based discount for each position in the ranking.\newline
- Constraints for the representation of underprivileged groups in the ranking. 
& - L-constraints. 
& - A ranking that maximizes the latent utility subject to the given constraints.
& - Pre. 
& - IIT-JEE.  
& / 
& / 
& / \\ \hline

\cite{mehrotra2021mitigating} 
& - Handle noisy protected attributes to improve fairness. 
& - Set of items, where each item has a utility $w_i \geq 0$.\newline
- A desired subset size $n$.\newline
- Protected attributes for each item (which may be noisy, missing, or imputed using proxy information).\newline
- Unbiased probabilistic information about the true protected attributes. 
& - LP approximation of denoising problem. 
& - A subset of $n$ items with the largest total utility that also satisfies the fairness constraints given the noisy protected attributes and the unbiased probabilistic information about the true protected attributes. 
& - Pre. 
& - US 2010 Census.\newline
- Income.\newline
- Synthetic.  
& - 'Risk difference'.
& - 'Utility ratio'. 
& - Yes. \\ \hline

\cite{asudeh2019designing} 
& - Improve fairness by choosing weights of the score function. 
& - $n$ items, each with several potentially relevant attributes.\newline
- Fairness constraints specifying a minimum bound on the number of selected members of a protected group at the $top-k$.\newline
- Scoring function that associates non-negative weights with item attributes and computes item scores. 
& - Satisfactory regions. 
& - A fair scoring function that satisfies the fairness constraint and is as close as possible to the user-specified scoring function in terms of attribute weights.\newline
- An ordered list of items sorted by their scores under the fair scoring function. \newline\newline
& - Pre and post. 
& - COMPAS.\newline
- US Department of Transportation.  
& / 
& / 
& Yes. \\ \hline

\multicolumn{10}{|c|}{\textbf{INFERENCE MODEL BASED}} \\ \hline
\cite{yang2020causal} 
& - Consider intersectionality in score based tasks and learn to rank tasks in a new way. 
& - Structural causal model.\newline
- Sensitive attributes.\newline
- $Y$ outcome variable used as a utility score in the ranking task.\newline
- Non-sensitive predictor variables.\newline
- Training dataset. 
& - Causal inference model. 
& - Score-based result: Counterfactually fair ranking of the dataset based on the utility score $Y$.\newline
- Learning to rank result: A learned model based on the counterfactual training data that can be used to rank the unmodified test data. 
& - Post. 
& - COMPAS.\newline
- MEPS.\newline
- Synthetic.  
& - Demographic parity.\newline
- Equal opportunity.\newline
- NDKL divergence.\newline
- IGF Ratio. 
& - $Y$ utility loss at $top-k$.\newline
- Average precision. 
& Yes. \\ \hline

\cite{mhasawade2021causal} 
& - Improve fairness by considering a causal multi-level fairness model. 
& - Set of variables associated with the system including individual-level and macro-level sensitive attributes.\newline
- Individual-level variables, macro-level variables.\newline
- An outcome of interest.\newline
- A causal graph consisting of V which accurately represents the data-generating process.
& - Causal inference model.\newline
- Multi-level path-specific fairness. 
& - A classifier that is fair with respect to both individual-level and macro-level sensitive attributes. 
& - Post. 
& - UCI Adult dataset.\newline
- Synthetic.  
& / 
& / 
& Yes. \\ \hline

\multicolumn{10}{|c|}{\textbf{METRICS BASED}} \\ \hline
\cite{ghosh2021fair} 
& - Highlight how demographic inference impacts fairness in rankings. 
& - Demographic information (either ground-truth or inferred) of individuals being ranked.\newline
- Error rates of demographic inference algorithms.\newline
- Fair rankings of individuals. 
& - DetConstSort. \newline
& - Evaluation of the impact of demographic inference errors on the fairness of the rankings. 
& - Post. 
& - Chess Rankings.\newline
- Crunchbase Entrepreneurs Ranking.\newline
- Equestrian Rankings.\newline
- Synthetic.  
& - Represen-\newline tation based: Skew; NDKL.\newline
- Attention-based: Attention per group; Attention bias ratio. 
& - NDCG.\newline
- Rank Change. 
& Yes. \\ \hline

\cite{lum2022biasing} 
& - Develop a statistically unbiased measure of group-wise performance disparities. 
& - Performance differences across socially or culturally relevant groups.\newline
- Statistical bias in current metrics used to measure group-wise performance disparities.
& - Double-corrected variance estimator. 
& - A statistically unbiased estimator for summarizing group performance disparities with associated uncertainty quantification. 
& - Post. 
& - Adult Income.\newline
- Synthetic.  
& - 'meta-metrics', showing their biases. 
& - Selection rate.\newline
- FPR.\newline
- TPR. 
& Yes. \\ \hline

\cite{lum2022biasing} 
& - Highlight the inadequacies of current evaluation methods in intersectionality. 
& - Demographic attributes for intersectional analysis. 
& - Introduction of new metrics. 
& - Comparisons between new and current metrics. 
& - Post. 
& - US 2010 Census.  
& - Max TPR difference.\newline
- Group rank. \newline
- Ranking correlation.
& / 
& / \\ \hline

\cite{ghosh2022faircanary} 
& - Solve the problem of drift in machine learning and artificial intelligence models, specifically with regards to ensuring fairness over time. 
& - Continuous model monitoring systems for detecting issues.
& - Quantile Demographic Drift.\newline
- FairCanary system. 
& - Identification of issues with models at deployment time and mitigations applied by developers. 
& - Post. 
& - Synthetic.  
& - Statistical Parity Difference.\newline
- Disparate Impact. 
& / 
& / \\ \hline

\cite{kang2022infofair} 
& - Ensure statistical parity among demographic groups formed by multiple sensitive attributes of interest. 
& - A set of sensitive attributes.\newline
- A set of data points that includes feature vectors, labels, and sensitive attribute values.\newline
- A learning algorithm. 
& - INFOFAIR method. 
& - Revised learning outcomes that mitigate bias and optimize classification accuracy.\newline
- Statistical parity on multiple sensitive attributes. 
& - Post. 
& - COMPAS.\newline
- Adult Income.  
& - Average statistical imparity.\newline
- Relative bias reduction. 
& - Micro/Macro F1. 
& Yes. \\ \hline

\cite{cachel2022mani} 
& - Combine the preferences of many rankers while ensuring fair treatment for candidates with multiple protected attributes. 
& - Multiple base rankings over candidates defined by multiple and multi-valued protected attributes.
& - MANI-RANK method. 
& - A consensus ranking that reflects the preferences of all rankers while ensuring fair treatment for candidates with multiple protected attributes. 
& - Post. 
& - Synthetic. 
& - Pairwise Disagreement loss.\newline
- Favored Pair Representation.\newline
- Attribute Rank Parity.\newline
- Intersectional Rank Parity. 
& - Kendall's Tau distance.
& Yes. \\ \hline


\caption{Synoptic table on intersectionality in rankings}
\label{tab:synoptic_table} 
\end{longtable}

\egroup
\end{landscape}
\newpage
\normalsize

\subsubsection{Discussion}
\label{subsec:intrank_table_results}

The systematic analysis of the studies presented in this section as well as a comparison via a synoptic table has highlighted the significant role of intersectionality in the context of fair rankings.

Our analysis demonstrates that intersectionality has several practical repercussions and can be implemented in actual 
ranking systems.
Indeed, the reviewed papers provided solid instances where intersectional considerations were effectively integrated into algorithmic processes.

Moreover, the significance of intersectionality in achieving truly equitable rankings has been vividly showcased.
While providing a certain level of equality, classical approaches to fairness often overlook intersecting identities and experiences. For example, in \cite{lum2022biasing} and \cite{wang2022towards} (Subsection \ref{subsec:intrank_meth_metr}), it is convincingly argued that traditional fairness metrics may need to catch up in capturing the complexities of multi-dimensional identities and their influence on outcomes. As a result, conventional methods can be considered inadequate when intersectionality is at stake.

One of the most crucial findings from our survey is that incorporating intersectionality in fair rankings does not necessarily lead to a significant loss in \emph{utility}. A notable example is \cite{mehrotra2021mitigating} (Subsection \ref{subsec:intrank_meth_constr}), where the authors achieved intersectional fairness without considerably compromising the algorithm's effectiveness.
This suggests that it is not just possible to introduce intersectionality in fair rankings, but it can be accomplished in most cases without causing significant drawbacks. Several methods are provided to balance potential losses, thereby ensuring the maintenance of intersectionality.

However, implementing intersectionality and, in general, fairness in rankings poses several challenges.
Among these, determining an acceptable threshold or balance that satisfies all stakeholders while maintaining the intended level of fairness is an extremely difficult task. Similarly, constructing a robust and universally acceptable ethical model, as discussed in \cite{yang2020causal} (Subsection \ref{subsec:intrank_meth_inf}), is another non-trivial issue.

\section{Conclusion}
\label{ch:conclusions}%

Rankings hold significant weight in multiple sectors, from recommendation systems to hiring processes. Despite the crucial role of these systems, we have identified consistent instances of bias that lead to discriminatory and prejudiced results.
In this respect, we have focused on ethical rankings and fairness in computer science. Addressing fairness and mitigating bias in algorithms, particularly regarding protected attributes such as race, gender, and socio-economic status, is critical.

Nevertheless, a significant gap in the current fairness-aware ranking methods lies in their tendency to focus on individual protected attributes, often ignoring the intertwined nature of these attributes that shape a person's identity.
The intersectionality perspective provides a more refined understanding of how bias and discrimination materialize in data-driven systems, particularly those impacting individuals belonging to multiple marginalized groups.

Our research has stressed the need to incorporate intersectionality into creating fairness-aware ranking algorithms. By considering multiple protected attributes together, the intersectional approach to fair rankings could achieve more equitable outcomes.
Our study offered a comparative analysis of existing literature on intersectional fair ranking in computer science through practical examples and provided a synoptic table to more easily compare the different methods.

We highlighted how fairness does not necessarily imply intersectionality, emphasizing the need for future research to explore and further develop intersectional approaches to fairness in ranking systems. More comprehensive techniques considering the intersection of multiple protected attributes could enable the development of ranking systems that more accurately reflect our diverse and intersectional society. Moreover, these approaches could contribute not only to more unbiased data-driven systems, but also to a more equitable representation of data in general, enhancing the integrity and reliability of such systems. 


\section*{Acknowledgments} 
Davide Martinenghi acknowledges support from the Italian PRIN project 2022XERWK9 ``S-PIC4CHU'' -- Semantics-based Provenance, Integrity, and Curation for Consistent, High-quality, and Unbiased data science

\bibliographystyle{unsrt}  
\bibliography{references}  

\end{document}